\documentclass{article}
\usepackage{graphicx}
\usepackage{amsmath,amssymb,amsfonts,graphics,setspace,tensor}
\usepackage{float}
\usepackage{slashed}
\usepackage[linktocpage=true,colorlinks,pdftex]{hyperref}
\usepackage{xcolor}
\colorlet{linkequation}{blue}
\usepackage{bm}
\usepackage{doi}
\hypersetup{colorlinks, citecolor=violet, filecolor=black, linkcolor=black, urlcolor=blue}
\usepackage{caption}
\usepackage{subcaption}
\title{About Fractional Calculus and its Applications in Physics}
\author{J.J.A. de Oliveira, C.F.L. Godinho\\
	Group of Theoretical Physics and Mathematical Physics,\\Department of Physics, Federal Rural University of Rio de Janeiro,\\
	Cx. Postal 23851, BR 465 Km 7, 23890-000 Seropédica - RJ,
	Brazil.\\
    	\texttt{fisica1jonjordan@gmail.com}, \texttt{crgodinho@gmail.com}}
\date{July 2025}

\begin{document}
\maketitle
\begin{abstract}
		\noindent
		Historically the fractional calculus concept works an extended idea based on the question asked by Guillaume de L'Hôpital to Gottfried Wilhelm Leibniz in 1695 about the notation ${d^nf}/{dx^n}$  for the derivative operator "What if $n=\frac{1}{2}$ ?"  To which Leibiniz replied : "This is an apparent paradox, from which useful consequences will be established".
		Our work revisits the unfolds who followed this questions with some classical definitions of fractional derivative operators and fractional integral.  We still  point out possible applications in areas such as Engineering, Physics, among others.  Among these definitions we will focus more on the Riemann-Liouville and Caputo definitions, however other definitions are also briefly commented.  In this work we begin with a historical inspection of the birth of the fractional calculus, parallels with the differential calculus and some of its developments are traced.  Always focusing on the definitions of Riemann-Liouville and Caputo, more commonly found in the bibliography of the area and more frequent in scientific works.  Some examples of its operability are presented, such as the direct calculation of constant function derivatives, polynomial function and exponential function.  Derivative operators and fractional integrals are defined as derivatives and noninteger-order integrals.  Our work revisits these two classical definitions of derivative operators and fractional integral and points out possible applications in areas such as Engineering, Physics, among others.  Our main goal is to address the feasibility of implementing this content in a degree form program in Physics, we strogly believe that this theme will aggregate a lot of content mainly because its multidisciplinary character.
		
		\textbf{Keywords:} Riemann-Liouville, Caputo, Fractional Calculus, FALVA.
	\end{abstract}
\newpage
\section{Introduction}
The ordinary Differential Calculus is a particular case of the Fractional Calculus, that was idealized more than 300 years ago.  The origin of the fractional calculus takes back to the birth of the theory of differential calculus, maybe based on fact that Leibniz have idealized the notation ${d^nf}/{dx^n}$ and the simple question made for L'Hôpital to Leibiniz in 1695 "What if $n=\frac{1}{2}$ ?" Historically many authors considered that problem, Johann Bernoulli and Leibniz used to mention derivatives of general order in private comunicaton.  The notation $d^{1/2}$ was used for John Wallis and Leibniz correspondence.  Obviously Leonard Euler gave attention to this problem and wrote about it in 1730 when suggested that if $n\in \mathbb{Z^{+}}$ and if $p=f(x)$ the ratio $d^{n}/dx^{n}$ must be always expressed algebraically. 
	
	Important to mention that several important names of mathematics have contributed to fractional calculus \cite{Miller & Ross}, S. F. Lacroix (1819) obtained the expression for the $n-$th derivative $d^{n}y/dx^{n} = \Gamma(m - 1)/\Gamma(m - n + 1) x^{m - n}$ and with it obtained the interesting result $d^{1/2}y/dx^{1/2} = 2\sqrt{x/\pi}$, P. S. Laplace (1820) proposed an integral definition for the fractional derivative, next was J.B.J. Fourier who wrote from his integral representation an interesting representation for the fractional derivative $d^{n}f(x)/dx^{n} = 1/{2\pi} \int^{\infty}_{-\infty}\int^{\infty}_{-\infty} f(\alpha)p^{n}cos [p(x - \alpha)+n\pi/2] d\alpha dp$. However for J.L Lagrange (1949) was attributed a indirect contribution, because the law of addition of indices, gave an indirect contribution when wrote the law of exponents for differential operators $d^{n}y/ dx^{n} \cdot d^{m}y / dx^{n} = d^{m + n}y/ dx^{m + n}$.  N. H. Abel (1823) considered the idea of fractional calculus for solving the tautochrone probem, Abel had developed a complete framework of what is today can be understood as fractional calculus, but unfortunately its inherent operational complexity prevented its application in a more large way.  After one decade of hibernation the fractional calculus was once more considered when J. Liouville published three memories whose his definitions were celebrated, his second definition is given by $	D^{\nu} x^{-a} = (-1)^{\nu} \Gamma(a + \nu)/{\Gamma(a)} x^{- a - \nu}$.  Even if posthumously, G.F.B. Riemann (1892) also contributed in during his student days with de definition of the fractional integral $D^{-\nu}f(x) = 1/ \Gamma(\nu) \int^{x}_{c} (x - t)^{\nu - 1} f(t) dt + \psi(x)$.
	
	The famous definition so called Riemann-Liouville definition was presented in a paper by N. Ya Sonin (1869) \cite{Sonin} where he based on a Cauchy's integral formula to reach an expression for differentiation with arbitrary index.  This definition today is one of the most famous and useful to study in a first moment coarse graining sistems with problems in areas such as anomalous difusion, rheology,  viscoelasticity, biological population models, robotics, and signal processing  maybe the most used version is given by.  The most recent and important contribution in the Fractional Calculus development was proposed by  M. Caputo (1967), where he re-defined the classical  Riemann-Liouville version of fractional derivative, $^{C}_{a}D^{\alpha}_{x} f(x) = 1/{\Gamma(n - \alpha)}\int^{x}_{a} (x - u)^{n-\alpha - 1}(df/du)^{n} du$.       
	
	As commented before, fractional calculus is a natural extension of ordinary differential calculus. Since the inception of the theory of differential and integral calculus, several ideas and applications about the calculation of non-integer order derivatives and integrals have been studied with many research groups around the world. Perhaps the subject would be more aptly called “integration and differentiation of arbitrary order.” Despite the work that has been done in this area, the application of fractional derivatives and integrals has been infrequent until recently. However, in recent years, advances in the theory of chaos and fractals revealed relationships with fractional derivatives and integrals, leading to renewed interest in this field.  The basic aspects of the theory of fractional calculus are outlined in.  As for the adoption of this concept in other scientific areas, several researchers have been inspired to examine this new possibility.  Some work has been carried out in the field of dynamical systems theory, but the proposed models and algorithms are still in the preliminary stage.  With these ideas in mind, this work introduces the fundamentals of fractional order calculus outand its applications focusing mainly in the Riemann-Liouville and Caputo definitions . This work is organized as follows. Section 2 outlines some aspects about the origins of fractional calculus and directs attention to the  Riemann-Liouville and Caputo derivatives, some important and useful properties are discussed, such as the relation between them. In the section 3 we present a more recent idea based on a variational unfold of the Riemann-Liouville integral named Fractional Actionlike Variational Approach (FALVA), some important examples of applications and areas where this approach is potentially usefull. Finally, section 4 presents a brief discussion where we consider the importance of this formalism, we still present some important areas of application and outlook for future perspectives.  
\section{About Fractional Calculus}
	\subsection{The Riemann-Liouville (RL) Fractional Integral Definition}
	
	As commented before we present some intrinsec aspects of the (RL) fractional integral .  It was also pointed that the earliest work that ultimately led to what is now called the Riemann-Liouville definition appears to be the paper by N. Ya. Sonin in 1869 \cite{Sonin} where he starting with the  Cauchy`s integral formula to achieve the differentiation with arbitrary index and  A. V. Letnikov extended Sonin's idea  in 1872 \cite{Let1,Let2}.  The main objective in both cases is to obtain the fractional derivatives idea by utilizing a closed contour. The starting point was the Cauchy's integral formula for integer order derivatives,  
	\begin{equation}
	\label{Cauchy integral}
	D^{n}f(z) = \dfrac{n!}{2\pi i}\int_{C} \dfrac{f(\zeta)}{(\zeta - z)^{n + 1}} d\zeta\,,
	\end{equation}
	the fractional version can be reached by replacing the factorial with Euler's Gamma function.  However, the extension to non-integer values $\alpha$ results in the problem that the integrand in (\ref{Cauchy integral}) does not have a pole anymore, instead contains a branching point and certainly an appropriate contour would require a branch cut which was not included in the work of Sonin and Letnikov yet. Laurent \cite{Laurent}, also considered Cauchy's integral formula with a contour given as an open circuit
	instead of a closed circuit proposed by Sonin and Letnikov, this led to the known definition of the RL fractional integral.

	\begin{align*}
	^{RL}_{c}D_{x}^{- \alpha} f(x) =  \dfrac{1}{\Gamma(\alpha)}\int^{x}_{c} (x - t)^{\alpha - 1} f(t) dt\,.
	\end{align*}
	Two distinct notations are usually considered for this kind of definition, both with the same meaning

	\begin{equation}
	_{a}I_{x}^{\alpha} [f(x)] = ^{RL}_{c}D_{x}^{- \alpha} f(x),
	\end{equation}
	henceforward we will adopt the first notation for the RL integral.	
	
	Now let us present some properties of the RL integral\footnote{$L_p(a,b)$ with $1\leq p\leq \infty$ is a set of those Lebesgue complex valued measurable function $f$},
	
	\begin{equation}
	P.1\,\,\,	_{a}I_{x}^{\alpha} (cf(x)) = c [_{a}I_{x}^{\alpha} (f(x))]\,,
	\end{equation}
	
	\begin{equation}
	P.2\,\,\,	_{a}I_{x}^{\alpha} (f(x) \pm g(x)) = _{a}I_{x}^{\alpha} (f(x)) \pm _{a}I_{x}^{\alpha} (g(x))\,,
	\end{equation}
	
	\begin{eqnarray}
	\label{RLDO} 
	P.3\,\,\,	_{a}I_{x}^{\alpha}(_{a}I_{x}^{\beta} f(x)) = _{a}I_{x}^{\alpha + \beta} (f(x)) \nonumber \\
	\alpha,\beta \geq 0\,,\,\,f(x)\in L_1[a,b]\,,
	\end{eqnarray}
	\begin{eqnarray}
	P.4\,\,\,\int_{a}^{b} {}_aI^{\alpha}_t f(x)g(x)dx&=&\int_{a}^{b}  f(x){}_tI^{\alpha}_bg(x)dx\,.
	\end{eqnarray}	
	
	Now we will prove (\ref{RLDO}) for considering that it is no a trivial consequence.  Let us start considering the RL integral definition
	
	\begin{equation}
	_{a}I_{x}^{\alpha} [f(x)] = \dfrac{1}{\Gamma(\alpha)} \int^{x}_{a} (x - t)^{\alpha - 1}f(t) dt \,,
	\end{equation}
	we then add another integral $I_{x}^{\beta}f(t)$ inside the first one with $\alpha>0, \beta>0 $ and $f$ continuous, then for all $x$ ,
	
	\begin{equation}
	_{a}I_{x}^{\alpha}[_{a}I_{x}^{\beta}f(x)] = \dfrac{1}{\Gamma(\alpha)} \int^{x}_{a} (x - t)^{\alpha - 1}[_{a}I_{x}^{\beta}f(t)] dt\,. 
	\end{equation}
	By definiton of the RL fractional integral and making use of Dirichilet's formula we rewrite the equation above as
	\begin{equation}
	_{a}I_{x}^{\alpha}[_{a}I_{x}^{\beta}f(x)] = \dfrac{1}{\Gamma(\alpha)} \int^{x}_{a} (x - t)^{\alpha - 1}\left[\dfrac{1}{\Gamma(\beta)} \int^{t}_{a} (t - \xi)^{\beta - 1}f(\xi) d\xi \right]\,, 
	\end{equation}
	and we can put the expression in the form
	\begin{equation}
	\label{Integral1}
	_{a}I_{x}^{\alpha}[_{a}I_{x}^{\beta}f(x)] = \dfrac{1}{\Gamma(\alpha)\Gamma{(\beta)}} \int^{x}_{a}\int^{t}_{a} (x - t)^{\alpha - 1}(t - \xi)^{\beta - 1}f(\xi) d\xi dt
	\end{equation}
	The expression (\ref{Integral1}) is valid for functions inside $[a,b]$, then using Fubini's theorem to compute the double integral, we can work with the double integral as an iterated integral:
	
	\begin{eqnarray}
	_{a}I_{x}^{\alpha}[_{a}I_{x}^{\beta}f(x)]&=& \dfrac{1}{\Gamma(\alpha)\Gamma{(\beta)}}\int^{x}_{a} f(\xi) d\xi \int^{x}_{\xi} (x - t)^{\alpha - 1}(t - \xi)^{\beta - 1} dt
	\end{eqnarray}
	
	Let us make a change of variable to simplify the whole expression, $ t = \xi + s(x - \xi) $, and consequently we have $ dt = (1 - s)d\xi $ and therefore
	
	\begin{eqnarray}
	_{a}I_{x}^{\alpha}[_{a}I_{x}^{\beta}f(x)] & = & \dfrac{1}{\Gamma(\alpha)\Gamma{(\beta)}} \int^{x}_{a} \int^{1}_{0} f(\xi)(x -(\xi + s(x - \xi))^{\alpha - 1}(\xi + s(x - \xi) -\xi)^{\beta -1}(1 - s)^{\alpha - 1}s^{\beta - 1} ds d\xi \nonumber\\
	\end{eqnarray}
	and we can write now
	
	\begin{equation}
	_{a}I_{x}^{\alpha}[_{a}I_{x}^{\beta}f(x)] = \dfrac{1}{\Gamma(\alpha)\Gamma{(\beta)}}\int^{x}_{a}\int^{1}_{0} f(\xi)(x - \xi)^{\alpha - \beta - 1}(1 - s)^{\alpha - 1} s^{\beta - 1} ds d\xi.
	\end{equation}
	
	\begin{equation}
	\label{beta}
	_{a}I_{x}^{\alpha}[_{a}I_{x}^{\beta}f(x)] = \dfrac{1}{\Gamma(\alpha)\Gamma{(\beta)}}\left(\int^{x}_{a} f(\xi)(x - \xi)^{\alpha - \beta - 1}d\xi\right)\left(\int^{1}_{0}(1 - s)^{\alpha - 1} s^{\beta - 1} ds \right).
	\end{equation}
	The expression (\ref{beta}) presents in the second factor the Euler's beta function
	
	\begin{equation}
	B(\alpha,\beta) = \int^{1}_{0}(1 - s)^{\alpha - 1} s^{\beta - 1} ds\,,
	\end{equation}
	
	\begin{equation}
	B(\alpha,\beta) = \dfrac{\Gamma(\alpha)\Gamma(\beta)}{\Gamma(\alpha)+ \Gamma(\beta)}\,.
	\end{equation}
	
	With this information we can write 
	
	\begin{equation}
	_{a}I_{x}^{\alpha}[_{a}I_{x}^{\beta}f(x)] = \dfrac{1}{\Gamma(\alpha)\Gamma{(\beta)}}\dfrac{\Gamma(\alpha)\Gamma(\beta)}{\Gamma(\alpha)+ \Gamma(\beta)}\left(\int^{x}_{a} f(\xi)(x - \xi)^{\alpha - \beta - 1}d\xi\right) 
	\end{equation}
	
	\begin{equation}
	_{a}I_{x}^{\alpha}[_{a}I_{x}^{\beta}f(x)] = \dfrac{1}{\Gamma(\alpha)+ \Gamma(\beta)}\int^{x}_{a} f(\xi)(x - \xi)^{\alpha - \beta - 1}d\xi\,.
	\end{equation}
	
	According with the RL integral definition we can see a connection with semigroups of operators. 
	
	\begin{equation}
	_{a}I_{x}^{\alpha}[_{a}I_{x}^{\beta}f(x)] = _{a}I_{x}^{\alpha + \beta}f(x)\,,
	\end{equation}
	
	and this complete the proof.

	\subsection{The Riemann-Liouville (RL) Fractional Derivative}
	
	Now, once we have understood the RL fractional integral definition, we can introduce the idea of fractional derivative.  The main idea of that operator is based on an inverse operator of the integral one in the sense

	\begin{equation}
	\dfrac{d}{dt} If(t) = f(t)\,.
	\end{equation}
	
	Now, we will consider the so called Riemann-Liouville derivative.  But first let us denote by $A C^{n}(\Omega)$ where $n=1,2,...$ and $\Omega$ is some interval, as an space of functions wich have continuous derivatives up to order $n-1$ on $\Omega$ with $f^{(n-1)}(x) \in A C^{n}(\Omega)$.  With this in mind, let us consider $f(x) \in A C^{n}[a,b]$ and  $ \alpha \in C $ with $ Re(\alpha) \geq 0 $ and $ \alpha \neq N $. The left and right fractional derivatives $ D_{a^{+}}^{\alpha} $, $ D_{a^{-}}^{\alpha} $ are defined by \cite{Sam,Teodoro et al}
	
	\begin{equation}
	\label{FD1}
	(D_{a^{+}}^{\alpha} f(x)) = \left(\dfrac{d}{dx}\right)^{n} (I_{a^{+}}^{n - \alpha} f(x))\,,   
	\end{equation}
	
	\begin{equation}
	\label{FD2}
	(D_{b^{-}}^{\alpha} f(x)) = \left(\dfrac{- d}{dx}\right)^{n} (I_{b^{-}}^{n - \alpha} f(x))\,. 
	\end{equation}
	
	Where $ [\alpha] $ is the real part of $ \alpha $,  $ Re(\alpha) $ with  $ n = [\alpha] + 1; x > a $ and  $ n = [\alpha] + 1; x > b $.  And from equations above we can understand that the arbitrary order derivative or fractional derivative, following RL arguments is an integer order derivative from an integral of arbitrary order, and as we defined before:

	\begin{equation}
	I_{a^{+}}^{n - \alpha} f(x) = \dfrac{1}{\Gamma(n - \alpha)} \int^{x}_{a} \dfrac{f(t)}{(x - t)^{n - \alpha + 1}} dt \,\,\,\,\,    x > a\,,
	\end{equation}
	
	\begin{equation}
	I_{b^{-}}^{n - \alpha} f(x) = \dfrac{1}{\Gamma(n - \alpha)} \int^{b}_{x} \dfrac{f(t)}{(x - t)^{n - \alpha + 1}} dt  \,\,\,\,\,     x < b\,.
	\end{equation}
	
	Then follows that
	
	\begin{equation}
	D_{a^{+}}^{\alpha} f(x) = \left(\dfrac{d}{dx}\right)^{n} \dfrac{1}{\Gamma(n - \alpha)} \int^{x}_{a} (x - t)^{n - \alpha + 1} dt \,\,\,\,\,    x < a\,,
	\end{equation}
	
	\begin{equation}
	D_{b^{-}}^{\alpha} f(x) = \left(-\dfrac{ d}{dx}\right)^{n}\dfrac{1}{\Gamma(n - \alpha)} \int^{b}_{x} (x - t)^{n - \alpha + 1} dt \,\,\,\,\,    x < b\,.
	\end{equation}
	
	Now as a particular case, let us consider a simple exercice considering $ \alpha=n \in \mathbb{N} $ then to recover the integer order derivative we have  \cite{Teodoro et al}
	
	\begin{equation}
	D_{a^{+}}^{n} f(x) = f^{n} (x)\,,
	\end{equation}
	
	\begin{equation}
	D_{b^{-}}^{n} f(x) = (-1)^{n} f^{n} (x)\,,  n \in \mathbb{N}\,.
	\end{equation}
	
	Where  $ f^{n}(x) $ is the n-fold derivative of $ f(x) $, and of course for $ \alpha = p \rightarrow 0 $, we have
	
	\begin{equation}
	\lim_{p\to 0} D_{a^{+}}^{p} f(x)=\lim_{p\to 0} D_{b^{-}}^{p} f(x) = f(x)
	\end{equation}
	
	The definitions (\ref{FD1}) and (\ref{FD2})  of the RL fractional derivatives  have contributed in the development of theory of fractional derivatives and integrals. This applications in pure mathematics presented a new class of probles under study as for instance fractional differential equations and new classes of functions.
	
	Recently there are many works where models are proposed in terms of fractional derivatives, models in areas as control processing, viscoelasticity, hereditary solid mechanics,signal processing ,anomalous diffusion, field theory and even dark matter.  Several works are based on the RL integral definition and its concept is used for a better understanding and description of such systems mainly that ones in the coarse graining regime.  However another important definiton is commonly considered in many works too.  In the next section we will introcuce the idea of the Caputo's fractional derivative.

	\subsection{Caputo Fractional Derivative}
	
	In this section we will present the Caputo's fractional derivative formulation and some paralel with the RL definition.  In a different way than RL formulation, Caputo's formulation is an integral of arbitrary order from an integer order derivative.  
	
	In his seminal paper \cite{Caputo}, Caputo have proposed a redefinition or a new definition of fractional derivative, by switching the order of the ordinary derivative with the fractional integral operator. By doing so, the Laplace transform of this new derivative depends on integer order initial conditions, differently from the initial conditions as the Riemann–Liouville fractional derivative, which works with fractional order conditions. Motivated by this concept, we address now the
	following definition. Let  $\alpha >0 $, $ n - 1 <  \mathbb{Re}(\alpha)\leq n $ then writing:
	\begin{equation}
	^{C}D^{\alpha}(t) = I^{n - \alpha}[D^{n} f(t)]\,,
	\end{equation}
	
	where $^{C}D^{\alpha}(t)$ refers to the Caputo's fractional derivative definition, $ D^{n} $ is a integer order derivative and $ I^{n - \alpha}$ is the RL fractional integral with $\alpha \in \mathbb{R_+} $ we have
	\begin{eqnarray}
	^{C}_{a}D^{\alpha}_{x} f(x) &=& _{a}I^{n - \alpha}_{x}D^{n}_{x}f(x) \,,\nonumber\\
	^{C}_{x}D^{\alpha}_{b} f(x) &=& _{x}I^{n - \alpha}_{b}D^{n}_{x}f(x) \,.\nonumber\\
	\end{eqnarray}
	Then, following the definitions presented above, the Caputo's fractional derivatives can be written as
	
	\begin{equation}
	^{C}_{a}D^{\alpha}_{x} f(x) = \dfrac{1}{\Gamma(n - \alpha)}\int^{x}_{a} \dfrac{f^{n}(u)}{(x - u)^{\alpha - n + 1}} du\,,
	\end{equation}
	
	\begin{equation}
	^{C}_{x}D^{\alpha}_{b} f(x) = \dfrac{(-1)^{n}}{\Gamma(n - \alpha)}\int^{b}_{x} \dfrac{f^{n}(u)}{(u - x)^{\alpha - n + 1}} du\,.
	\end{equation}
	Both equations are valid for $ n = [\alpha] + 1 $ , $ \alpha \in \mathbb{Re}^{\ast}_{+} $ , $ a \in \mathbb{Re} $.  As a direct consequence of the definition we clearly see that are nonlocal operators too because both, left
	[$^{C}_{a}D^{\alpha}_{x} f(x)$] and right [$^{C}_{x}D^{\alpha}_{b} f(x)$] operators will depend on its values both left and right of $ x $ ($ a\leq u \leq x $, $ x \leq u \leq b $).
	
	An interesting relation can be done between the Caputo and RL fractional derivatives 
	\begin{equation}
	{}_aD^{n+a}_x f(x)={}_a^CD^{n+a}f(x)+\sum_{k=0}^n\frac{f^{k}(0)x^{k-n-\alpha}}{\Gamma(k-\alpha-n+1)}\,.
	\end{equation}
	The fractional calculus is good tool for describing with more accuracy dissipative and coarse grained systems, its nonlocal nature confers indeed some different consequences such as memory effect.  In many works are considered the order of fractionality beteween zero and one, because we can infer analysis more close with the usual behaviour testing what sometimes is called weak fractionality $\alpha \approx 1$, once more we highlight that this is an interesting strategy when are are facing viscous and dissipative systems. 
	\section{FALVA}
	In this section we present a variational approach usually called Fractional Actionlike Variation Approach, or merely FALVA \cite{El Nabulsi & Delfim Torres}.	
	
	The main idea is based on the action principle inherent in the RL fractional integral.  Considering first the one-dimensional case where we consider a smooth n-dimensional manifold $M$ as a configuration space and denote by $L:TMM \times \mathbb{R} \times \mathbb{R}$ the smooth Lagrangian function. For any smooth path $q:[a,b]\rightarrow M$ satisfying ﬁxed boundary conditions $q(a) = q_a$ and $q(b) = q_b$, we then address  the idea of a  fractional action integral given by
	\begin{equation}
	\label{Falva1}
	S^{\alpha}[q](t) = \dfrac{1}{\Gamma(\alpha)}\int^{x}_{a} L(\dot{q}(\tau), q(\tau), \tau)(t - \tau)^{\alpha - 1} d\tau    \,,
	\end{equation}
	here $\dot{q} = \dfrac{dq}{d\tau}$ is the time derivative , $\alpha$ is the fractional order $(0 < \alpha < 1)$, $L(\dot{q}(\tau), q(\tau), \tau)$ is the lagrangean, $t\in [a,t]$ is the intrinsic time and $ \tau \in [a,b]$ is the oberserver time.  Using this fractional action to find the stationary point, we then have
	\begin{equation}
	\delta(S^{\alpha}) f(t) =  \dfrac{1}{\Gamma(\alpha)}\int^{t}_{a} (t - \tau)^{\alpha - 1} \delta(L(q(\tau), \dot{q}(\tau), \tau)) d\tau     
	\end{equation}
	and by the traditional methods of differentiating under signal of integral
	
	\begin{equation}
	\delta L(q(\tau), \dot{q}(\tau), \tau) = \sum_{i=1}^n\left[\dfrac{\partial L}{\partial q_i} \delta q_i + \dfrac{\partial L}{\partial \dot{q_i}} \delta \dot{q_i} \right]+ \dfrac{\partial L}{\partial \tau} d\tau\,,\,\,\,\,\,i=1,...,n
	\end{equation}
	and therefore imposing the stationary condition,
	
	\begin{equation}
	\label{FALVA2}
	\delta(_{a}I_{t}^{\alpha}) f(t) =  \dfrac{1}{\Gamma(\alpha)}\int^{t}_{a}\sum_{i=1}^n\left[\dfrac{\partial L}{\partial q} \delta q + \dfrac{\partial L}{\partial \dot{q}} \delta \dot{q}\right] (t - \tau)^{\alpha - 1}\delta\tau=0\,.
	\end{equation}
	The boundary conditions on the configuration space for $\delta q$ at points $a$ and $t$ are $q(a) = q(t) = 0$ and therefore using integration by parts we find that
	
	\begin{equation}
	\int^{a}_{t} \dfrac{\partial L} {\partial \dot{q}} \dfrac{d}{d\tau} \delta q (t - \tau)^{\alpha - 1} d\tau = -\int^{t}_{a} \dfrac{d}{d\tau} \left[\dfrac{\partial L}{\partial \dot{q}}(t - \tau)^{\alpha - 1}\right] \delta q d\tau\,.
	\end{equation}
	After integrate by parts and algebraic manipulations, we can rewrite (\ref{FALVA2}) as  
	\begin{equation}
	_{t}I_{a}^{\alpha} f(t) = \dfrac{1}{\Gamma(\alpha)} \int^{t}_{0} \sum_{i=1}^n\left[\dfrac{\partial L}{\partial q} - \dfrac{d}{d\tau}\left(\dfrac{\partial L}{\partial \dot{q}}\right) - \left(\dfrac{\alpha - 1}{t - \tau}\right)\dfrac{\partial L}{\partial \dot{q}}\right] (t - \tau)^{\alpha - 1} \delta q d \tau\,
	\end{equation}
	and since the generalized coordinates $q_i(\tau)$ are independent in the configuration space, and considering the fundamental lemma of the calculus of variations we are authorized to write that;
	\begin{equation}
	\sum_{i=1}^n\left[\dfrac{\partial L}{\partial q_i} - \dfrac{d}{d\tau}\left(\dfrac{\partial L}{\partial \dot{q_i}}\right) - \left(\dfrac{\alpha - 1}{t - \tau}\right)\dfrac{\partial L}{\partial \dot{q_i}}\right]= 0\,,
	\end{equation}
	and that is the fractional RL version of the Euler-Lagrange equations for some $t \in [a,b]$.  The meaning of the factor $\left(\dfrac{\alpha - 1}{t - \tau}\right)$ can be understood when we consider frictional forces and, particularly, when this force is proportional to the velocity and with components $j$ given by $F_{fj}=-k_jv_j\,,\,\,\,\,j=x,y,z$ or $j=1,2,3$. This formulation was idealized by John William Strutt as known as Lord Rayleigh and the Rayleigh's dissipation function $\mathcal{F}$ is defined as \cite{Goldstein}
	
	\begin{equation}
	\mathcal{F}={1\over2} \sum_{i}^n\sum_{j}^3 k_jv^2_{ij} \,.
	\end{equation}
	and as a direct unfolding,
	\begin{equation}
	F_{fj}=- \dfrac{\partial \mathcal{F} }{\partial v_j}   \,.
	\end{equation}
	The nature of this dissipative force allows us to write the usual Euler-Lagrange equation in the form
	\begin{equation}
	\dfrac{\partial L}{\partial q} - \dfrac{d}{d\tau}\left(\dfrac{\partial L}{\partial \dot{q}}\right) + \left(\dfrac{\partial \mathcal{F}}{\partial \dot{q}_{i}}\right)  = 0    
	\end{equation}
	and it is very clear that the sector where this force appears, is the same sector of the fractional contribution $\mathcal{F} = \dfrac{\alpha - 1}{t - \tau}$.
	\section{Conclusion}
	We have presented a brief review of fractional calculus motivated by its application in many areas of science.  The functional (\ref{Falva1}) is indeed very adopted today for studying many problems such as dynamical economic analysis, coarse graining systems with dissipation.  Some of these dissipative systems are for instance given by complex fluids, these ones can be understood as fluids that despite their continuum appearance at macroscopic scales have a structure at mesoscopic scales.  We can think of macromolecules like granular gases, bubbles, droplets and shaving scream as examples of such structures.  We still have many other fluids of technological importance, such as lubricants, paints, surfactant-aided oil recovery fluids, liquid crystals, plastics, shampoos, etc., are complex fluids. What makes these fluids 'complex' is the conspicuous coupling between the mesoscopic structures of the fluids and their flow properties.  Other area where recently has seem to be a good laboratory of using fractional approach is dark matter \cite{GODINHO}, where a nonlocal extension of the mimetic dark matter model based on the FALVA implementation of fractional calculus is proposed to study certain dark matter properties.  We still find very recent contributions with fractional differential equations in cosmology because of its memory effect \cite{Bayron}.  Considering these facts we strongly believe that the fractional formalism is today a real and consistent area of research.

\end{document}